\documentclass[journal]{vgtc}                     %

\onlineid{0}

\vgtccategory{Research}

\usepackage{inline_code}
\usepackage{enumitem}
\title{Agentic Authoring of Interactive Multiview Visualizations in Genomics}

\author{%
  \authororcid{Astrid van den Brandt}{0000-0002-3676-1341},
      \authororcid{Kiroong Choe}{0000-0002-6084-2539},
        \authororcid{Sehi L'Yi}{0000-0001-7720-2848},
            \authororcid{Devin Lange}{0000-0002-3467-0294},
  and 
  \authororcid{Nils Gehlenborg}{0000-0003-0327-8297}
}

\authorfooter{
  \item
  	Astrid van den Brandt, Sehi L'Yi, Devin Lange, and Nils Gehlenborg are with Harvard Medical School.
  	E-mail: {astrid\_vandenbrandt | sehi\_lyi | devin\_lange | nils}@hms.harvard.edu
  \item
  	Kiroong Choe is with Boston College
  	E-mail: choekir@bc.edu.

}

\abstract{The diverse types of genomics data, scientific questions, and resulting analysis tasks typically require highly specialized visualizations. Users often need to customize or author entirely new visualizations tailored to their specific data and analysis tasks. Although many visualization tools are available, they are typically either limited in the degree of customization they support or require extensive learning or even programming effort to use effectively. Moreover, even when a tool is sufficiently expressive, users may lack the visualization expertise to produce effective designs.
Agentic and large language model (LLM)-based approaches are increasingly used to assist in complex scientific and engineering tasks, including data visualization. The use of natural language through conversational interfaces is a promising avenue for democratizing the authoring of complex visualizations. In the context of genomics, these approaches face additional challenges: genomics visualizations typically integrate heterogeneous data types and are composed of multiple linked interactive views.
These challenges motivate the exploration of more structured LLM-based schemes. We first characterize where vanilla LLM generation succeeds and fails for genomics visualization, identifying eight quality dimensions. We then compare six schemes---direct generation, a fixed pipeline, and four agentic configurations varying in the number of specialist agents and the presence of a reviewer---across 159 cases spanning three levels of query ambiguity and specification complexity. All schemes use the Gosling visualization grammar as structured output. Our results show that agentic iteration substantially improves perceived quality over both baselines, while we did not observe additional benefits from more complex agent architectures. We discuss implications for the design of agentic systems for domain-specific visualization authoring.
All supplemental materials are available at \href{https://doi.org/10.17605/OSF.IO/UQE83}
{\texttt{osf.io/uqe83}}.
}


\teaser{
  \centering
  \includegraphics[width=\linewidth, alt={}]{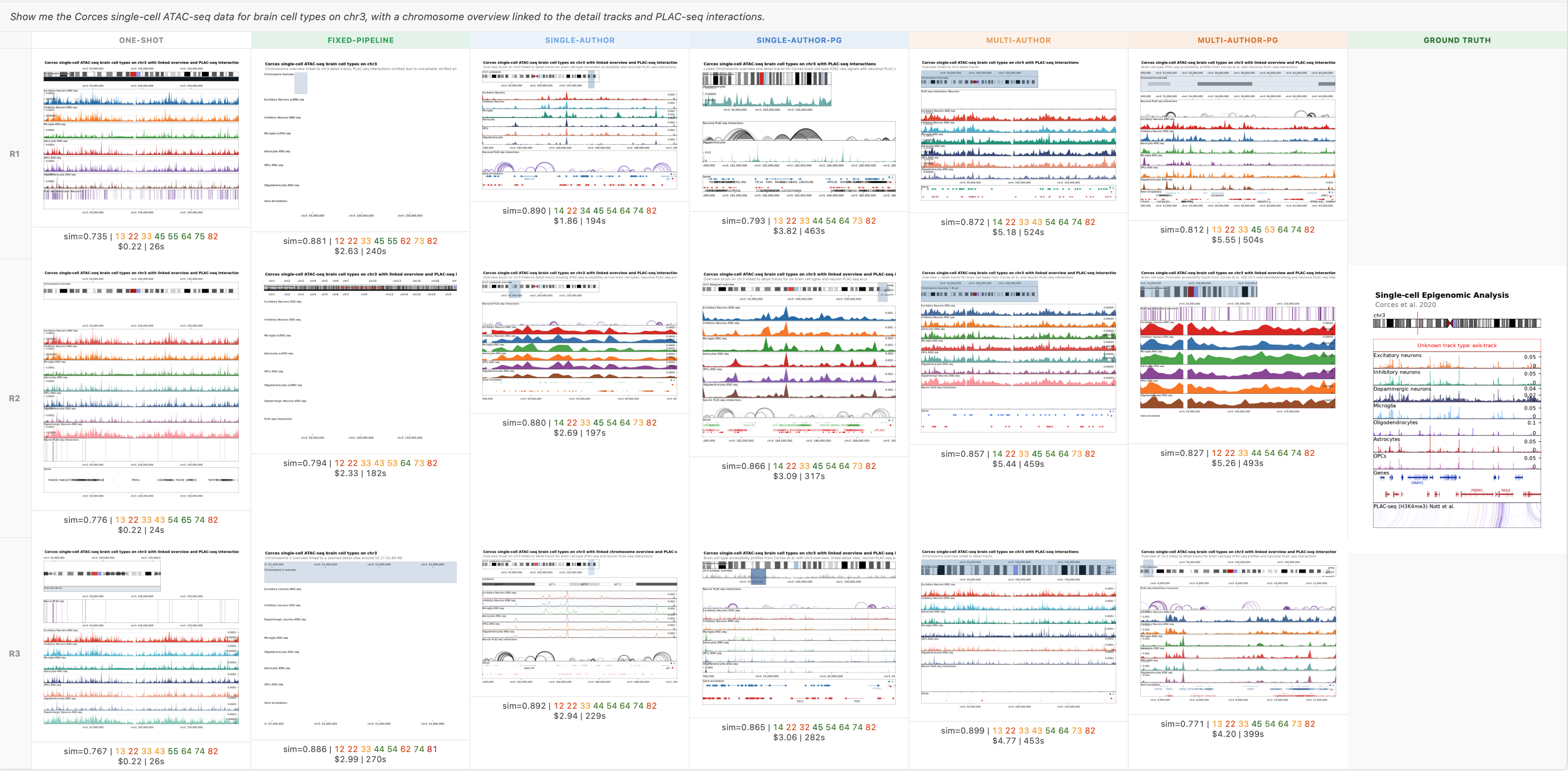}
  \caption{%
    Outputs from all six authoring schemes (columns: one-shot, fixed-pipeline, single-author, single-author-pg, multi-author, multi-author-pg) across three independent runs (rows: R1, R2, R3) for a complex L3 specification under the ambiguous S2 scenario using a single-cell epigenomics dataset~\cite{corcesSinglecellEpigenomicAnalyses2020}. %
  }
  \label{fig:teaser}
}

\renewcommand{\manuscriptnotetxt}{This is the authors' preprint version of this paper. License: CC-By Attribution 4.0 International.}

\graphicspath{{figs/}{figures/}{pictures/}{images/}{./}} %

\usepackage{tabu}                      %
\usepackage{booktabs}                  %
\usepackage{lipsum}                    %
\usepackage{mwe}                       %
\usepackage{ccicons}                   %
\usepackage{listings}
\lstdefinestyle{json}{
  basicstyle=\ttfamily\tiny,
  breaklines=true,
  backgroundcolor=\color{gray!8},
  frame=none,
  xleftmargin=1em,
  xrightmargin=1em,
  showstringspaces=false,
  string=[s]{"}{"},
  stringstyle=\color{jsonstring},
  escapeinside={(*@}{@*)},
  literate=
     {:}{{{\color{jsonpunct}{:}}}}{1}
     {,}{{{\color{jsonpunct}{,}}}}{1}
     {\{}{{{\color{jsonpunct}{\{}}}}{1}
     {\}}{{{\color{jsonpunct}{\}}}}}{1}
     {[}{{{\color{jsonpunct}{[}}}}{1}
     {]}{{{\color{jsonpunct}{]}}}}{1}
     {0}{{{\color{jsonstring}{0}}}}{1}
     {1}{{{\color{jsonstring}{1}}}}{1}
     {2}{{{\color{jsonstring}{2}}}}{1}
     {3}{{{\color{jsonstring}{3}}}}{1}
     {4}{{{\color{jsonstring}{4}}}}{1}
     {5}{{{\color{jsonstring}{5}}}}{1}
     {6}{{{\color{jsonstring}{6}}}}{1}
     {7}{{{\color{jsonstring}{7}}}}{1}
     {8}{{{\color{jsonstring}{8}}}}{1}
     {9}{{{\color{jsonstring}{9}}}}{1},
}

\usepackage{mathptmx}                  %

\usepackage{xcolor}

\newcommand{\csq}[1]{\textcolor{#1}{\rule{0.65em}{0.65em}}}
\definecolor{cQDA}{HTML}{0076BD}
\definecolor{cVEC}{HTML}{FF9600}
\definecolor{cLS}{HTML}{00A46C}
\definecolor{cShared}{HTML}{7F7F7F}
\definecolor{cPG}{HTML}{E66EAB}

\definecolor{todoBlue}{RGB}{0, 102, 204}

\definecolor{jsonkey}{RGB}{0, 80, 180}
\definecolor{jsonstring}{RGB}{0, 130, 60}
\definecolor{jsonnumber}{RGB}{180, 60, 0}
\definecolor{jsonpunct}{RGB}{100, 100, 100}
\definecolor{importantRed}{RGB}{204, 0, 0}
\definecolor{noteGreen}{RGB}{0, 140, 70}
\definecolor{warnOrange}{RGB}{210, 100, 0}
\definecolor{reviseGrey}{RGB}{120, 120, 120}

\begin{document}

\firstsection{Introduction}
\label{sec:introduction}

\maketitle
\makeatletter
\renewcommand{\@oddfoot}{\hfil\textrm{\thepage}\hfil}
\renewcommand{\@evenfoot}{\hfil\textrm{\thepage}\hfil}
\makeatother

Genomics research spans diverse data types, scientific questions, and analysis tasks, each typically demanding highly specialized visualizations~\cite{nusratTasksTechniquesTools2019}. As a consequence, the design space for genomics data visualization is vast; %
users often need to create or customize entirely new visualizations tailored to their specific data and analysis tasks.
Although many visualization tools are available, they are typically either limited in the degree of customization they support or require extensive learning or even programming effort to use effectively. This creates a lock-in effect, where users are constrained by the affordances of a single tool. Moreover, even when a tool is sufficiently expressive, users may lack the visualization expertise to produce effective designs~\cite{vandenbrandtUnderstandingVisualizationAuthoring2025}.

Large language model (LLM)-based approaches are increasingly used to assist users in complex scientific and engineering tasks. In the context of visualization, they have been applied to a growing range of tasks, most prominently for automatic visualization generation from natural language~\cite{maddiganChat2VISGeneratingData2023, dibiaLIDAToolAutomatic2023, goswamiPlotGenMultiAgentLLMbased2025a, lyiLearnableExpressiveVisualization2025, liPrompt4VisPromptingLarge2024, vazquezAreLLMsReady2024b, ribalta-albadoEvaluatingLLMsAbilities2026b, ponochevnyiTalkMeIt2026a, ouyangNVAGENTAutomatedData, zhangAuraGenomeLLMPoweredFramework2025, chenInterChatEnhancingGenerative2025, langeYACBridgingNatural2025, hostnikVegaChatRobustFramework2026, chenCoDAAGENTICSYSTEMS2026}, but also recommendation~\cite{wangDracoGPTExtractingVisualization2025, kimHowGoodChatGPT2025, ahnUnderstandingWhyChatGPT2025}, evaluation and visualization literacy~\cite{dasChartsofThoughtEnhancingLLM2025, dongProbingVisualizationLiteracy2025}, and design studies~\cite{ruanQualitativeStudyLLMassisted2025}. The use of natural language through conversational interfaces is a particularly promising avenue for democratizing complex visualization authoring, lowering the barrier for users who lack programming or visualization expertise.

While these models show capability for generic statistical visualizations, they remain limited when applied to complex, domain-specific data~\cite{zhangAuraGenomeLLMPoweredFramework2025, gaoFineTunedLargeLanguage2025, dasChartsofThoughtEnhancingLLM2025}. Genomics visualizations pose a particular challenge; they typically integrate heterogeneous data types and are composed of multiple linked, interactive views that support multiscale reasoning. Beyond these domain-specific visualization challenges, effective use of LLMs requires well-constructed prompts, which is itself a non-trivial, trial-and-error process. %
While LLMs show promise for specific, well-scoped genomics visualizations~\cite{zhangAuraGenomeLLMPoweredFramework2025}, the capabilities and failure modes when authoring the full breadth of complex genomics visualizations are not yet well understood.

Several strategies have been explored to adapt LLMs to specific tasks or domains, including supervised fine-tuning (SFT), prompt engineering, and retrieval-augmented generation (RAG). These can yield meaningful improvements, but each has its limits: fine-tuning is costly and data-hungry, prompt engineering can be ad hoc and brittle, and RAG depends heavily on quality and coverage. Moreover, LLM outputs are inherently stochastic and require steering.
Recently, there has been a shift from using a single general-purpose LLM toward agent-based systems, where multiple specialized agents collaborate, access tools, critique, and iteratively refine outputs~\cite{gaoDemocratizingAIScientists2025}. These so-called AI scientist systems offer several advantages: they avoid costly retraining, naturally support decomposition of complex tasks into smaller sub-tasks, and because agents operate within a defined scope and role, logging inputs and outputs in these more restricted contexts can improve provenance tracking, a known challenge in LLM-based systems~\cite{hongDataHasEntered2025}. 

Single and multi-agent approaches have been proposed for standard visualization~\cite{ouyangNVAGENTAutomatedData, luMultiVisAgentMultiAgentFramework2026, chenCoDAAGENTICSYSTEMS2026, goswamiPlotGenMultiAgentLLMbased2025a, yangMatPlotAgentMethodEvaluation2024} and for specific cases such as circular genome visualizations~\cite{zhangAuraGenomeLLMPoweredFramework2025}. 
While these works show the potential of multi-agent schemes for visualization authoring, it is often unclear why a particular set of agents was chosen or what alternatives were considered~\cite{leeAgenticDrawerAdvisor2025}, making it difficult to know how these designs would generalize to other authoring tasks. Existing approaches often struggle with complex, heterogeneous data (e.g., multiple large files and metadata~\cite{chenCoDAAGENTICSYSTEMS2026}), and they focus mostly on general-purpose visualizations or narrowly scoped domain-specific cases (e.g., AuraGenome~\cite{zhangAuraGenomeLLMPoweredFramework2025}), leaving broader genomics questions largely underexplored.
Finally, user interactions are rarely taken into consideration, yet they are likely essential for authoring complex, multiview visualizations.
Human input is needed for resolving ambiguity and incorporating domain-specific expertise into visualization design.
There is also no consensus on how to evaluate visualization authoring quality, with existing works using different metrics, making comparison across schemes difficult.

In this work, we explore how agent-based LLM schemes can support the construction of complex genomics visualizations with the Gosling visualization grammar~\cite{lyiGoslingGrammarbasedToolkit2022}. We begin by characterizing where a vanilla LLM succeeds and fails when generating Gosling specifications, identifying eight recurring quality dimensions. We then ask whether agentic iteration improves upon simpler approaches, and whether further architectural choices within agentic schemes matter, by comparing six schemes: direct generation, a hand-crafted fixed pipeline, and four agentic configurations varying in the number of specialist agents and the presence of a reviewer. To evaluate these schemes, we construct 159 cases from 53 ground-truth specifications crossed with three levels of query ambiguity, and assess output quality through both structural similarity to the reference and perceived quality scored by an LLM judge. The main contributions are:

\begin{itemize}

\item A characterization of LLM capabilities and failure modes for genomics visualization authoring along eight quality dimensions.

\item An evaluation framework that uses a user proxy agent to simulate realistic query ambiguity, enabling comparison of authoring schemes beyond fully specified benchmarks.

\item Findings from a comparative evaluation of six schemes on the effects of agentic iteration, agent architecture, specification complexity, and evaluation metrics.

\end{itemize}

Based on these findings, we discuss how specialized grammars and flexible agents can complement each other, and what this means for the design and evaluation of agentic visualization authoring systems.

\section{Background and Related Work}
\label{sec:background_relatedwork}

\subsection{Genomics Data Visualization}
\begin{figure*}[!t]
    \centering
    \includegraphics[width=1\textwidth, alt={}]{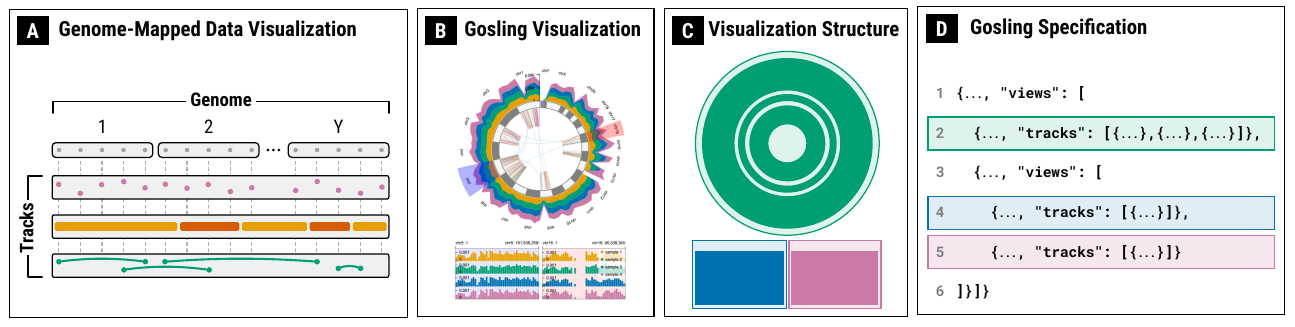}
    \caption{%
   The background of the genome-mapped data visualization and Gosling. (A) Different types of information can be mapped to a genome, which consists of chromosomes. Each data visualized is referred to as a track. (B--D) The Gosling visualization (B) and its specification (D), consisting of three views and five tracks (C), are shown.
    }
    \label{fig:background-gosling}
\end{figure*}

The genome is organized into chromosomes, each a long sequence of nucleotides (A, T, C, G) that encodes genetic information. Genomic datasets capture diverse signals along these sequences.
Genomic data visualizations organize these datasets into tracks, which are composed into multi-view layouts for comparative analysis~\cite{lyiGoslingGrammarbasedToolkit2022, lyiMultiViewDesignPatterns2023} (see~\cref{fig:background-gosling}). Common tasks include identifying patterns across chromosomal regions, comparing signals between samples, and navigating across scales from whole-genome overviews to single-nucleotide detail~\cite{nusratTasksTechniquesTools2019}. Detailed nomenclature for genomic visualization types, tracks, and tasks can be found in Nusrat et al.~\cite{nusratTasksTechniquesTools2019}.

Genomics data visualizations typically deal with domain-specific file formats such as BigWig (continuous signals), BED (genomic intervals), BAM (aligned sequence fragments), VCF (variants), and Hi-C matrices (chromatin contacts). These datasets can typically be browsed in genome browsers and other specialized tools (e.g., IGV~\cite{thorvaldsdottir2013integrative}, JBrowse~\cite{buels2016jbrowse}, MizBee~\cite{meyer2009mizbee})~\cite{nusratTasksTechniquesTools2019}. More recently, Gosling introduced a declarative grammar that specifies genomic visualizations as JSON specifications, enabling declarative authoring of tracks, views, and interactions such as brushing and coordinated zoom. Building on this grammar, systems like GenoREC~\cite{pandeyGenoRECRecommendationSystem2023a} provide rule-based recommendation for genomic visualization design, altGosling~\cite{smits2024altgosling} addresses accessibility, Blace~\cite{lyiLearnableExpressiveVisualization2025} proposes blended user interfaces, and Geranium~\cite{nguyenGeraniumMultimodalRetrieval2025} explores example search for authoring.

\subsection{Natural Language for Visualization Authoring}
The ability to author visualizations through natural language (NL2Vis) rather than code has been a longstanding goal in data visualization. The field has evolved from early rule-based and template-matching systems toward LLM-based approaches that can handle more diverse and underspecified queries. Various strategies for adapting LLMs to visualization authoring have been explored, from prompt engineering and fine-tuning to multi-step pipelines and agentic architectures. Systems such as LIDA~\cite{dibiaLIDAToolAutomatic2023}, Chat2Vis~\cite{maddiganChat2VISGeneratingData2023}, and ChartGPT~\cite{tianChartGPTLeveragingLLMs2025} demonstrate that LLMs can generate visualizations from natural language, typically targeting single-view, single-dataset scenarios. More recent work moves beyond single-call generation. For example, VegaChat~\cite{hostnikVegaChatRobustFramework2026} uses few-shot prompted LLMs within a multi-turn pipeline that allows follow-up messages for chart refinements, though evaluation remains focused on single-view cases. 

While these advances all broadened the capabilities of NL2Vis, most existing systems and benchmarks focus on standard chart types. Available datasets---such as NLVcorpus~\cite{srinivasanCollectingCharacterizingNatural2021}, MatPlotBench~\cite{yangMatPlotAgentMethodEvaluation2024}, ChartLLM~\cite{koNaturalLanguageDataset2024}, MultiVis-Bench~\cite{luMultiVisAgentMultiAgentFramework2026}, and VisEval~\cite{chenVisEvalBenchmarkData2025}---primarily contribute collections of natural language queries with single-view visualizations of generic datasets. DQVis~\cite{langeDQVisDatasetNatural2025} and GQVis~\cite{waltersGQVisDatasetGenomics2025} have begun to address biomedical data-question and genomics-specific queries respectively, but evaluation of complex, multi-view genomic visualizations remains largely unexplored. Whether LLMs can generate domain-specific visualizations and operate on specialized grammars like Gosling remains an open question. Such grammars involve multi-track composition, coordinated interactions, and genomic data formats. Our work addresses this gap with both an evaluation framework and an agentic authoring system tailored to this complexity.

\subsection{AI Agents for Visualization}
Recently, AI agentic architectures have emerged as a distinct category of NL2Vis systems, with the potential to handle more complex scenarios through task decomposition and iterative refinement, though often at increased cost in latency and token usage~\cite{hostnikVegaChatRobustFramework2026}. 
Several systems have explored different perspectives in this design space. nvAgent~\cite{ouyangNVAGENTAutomatedData} adopts a \emph{divide-and-conquer} paradigm with three specialized agents (processor, composer, validator) and introduces a Visualization Query Language (VQL) as an intermediate representation bridging natural language and visualization code. PlotGen~\cite{goswamiPlotGenMultiAgentLLMbased2025a} uses five agents running sequentially with self-reflection loops, placing it closer to a structured LLM workflow than a fully agentic system. MatPlotAgent~\cite{yangMatPlotAgentMethodEvaluation2024} and Drawer-Advisor~\cite{leeAgenticDrawerAdvisor2025} similarly explore agent-based generation with feedback mechanisms. CoDA~\cite{chenCoDAAGENTICSYSTEMS2026} introduces a self-evolving pipeline where agents specialize in understanding, planning, generating, and reflecting, using metadata schemas and statistics to circumvent context window limits.

Most relevant to our work are systems targeting interactive, multi-view or domain-specific visualization. MultiVis-Agent~\cite{luMultiVisAgentMultiAgentFramework2026} coordinates three sub-agents through a central orchestrator that performs dynamic reasoning and tool selection, while the sub-agents themselves act more as specialized tools, a common pattern that favors reliability over flexibility. AuraGenome~\cite{zhangAuraGenomeLLMPoweredFramework2025} is, to our knowledge, the only prior system targeting genomics visualization with an agentic approach. It employs seven specialized LLM agents for intent recognition, data parsing, layout recommendation, code generation, and refinement. While AuraGenome shares our motivation of supporting complex domain specific authoring, it is more narrowly scoped to single view, circular layout charts.

A common thread across these systems is that architectural choices, such as how many agents, how they communicate, whether feedback is visual or spec-level, are often asserted rather than empirically compared. MultiVis-Agent claims benefits of its centralized architecture but does not compare against alternatives. PlotGen's sequential ordering of feedback agents (numeric, lexical, visual) is mainly textually justified. Our work contributes to this space by systematically comparing authoring schemes, from one-shot to multi-agent with review, and grounded in empirically identified quality dimensions.

\section{Why Not Vanilla LLM Generation?}
\label{sec:compare-vanilla-llm}

It is poorly understood how well LLMs generalize to the full breadth of complex genomics visualizations~\cite{zhangAuraGenomeLLMPoweredFramework2025}. To characterize the capabilities and failure modes, we evaluate a vanilla one-shot generation as baseline. This is a single LLM call that receives a user query, a data catalog, the raw Gosling.js documentation, and outputs a complete Gosling specification or "Gosling spec" for short. The LLM is instructed to use dataset URLs and field names exactly as given in the catalog. The catalog includes metadata, field statistics, and sample rows, giving the model a fair shot at succeeding. The documentation and metadata furthermore mirror how a non-expert user would typically approach authoring a Gosling visualization. We use \textsc{gpt-5.4-2026-03-05} as the underlying model.

\subsection{Dataset}
The evaluation dataset combines examples from multiple data sources, organized by scenario (S1--S3) and complexity level (L1--L3). Examples are derived from Gosling's example gallery and Geranium~\cite{nguyenGeraniumMultimodalRetrieval2025}.
Each example pairs a natural language query with a ground truth Gosling spec for comparison. A total of 53 unique specs (L1: 20, L2: 22, L3: 11) are instantiated across all three scenarios, yielding 159 examples. An example entry is shown in Fig.~\ref{fig:dataset-example}.

\newsavebox{\datasetcodebox}
\begin{lrbox}{\datasetcodebox}\begin{minipage}{0.54\columnwidth}
\begin{lstlisting}[style=json]
{
  (*@\textcolor{jsonkey}{"query"}@*): "Show me my H3K27ac ChIP-seq data across the whole genome.",
  (*@\textcolor{jsonkey}{"Q"}@*): "Create a bar chart of peak values for sample 1 across the whole hg38 genome from H3K27ac-resgen.multivec, with each bar spanning its bin size.",
  (*@\textcolor{jsonkey}{"task"}@*): "SUMMARIZE",
  (*@\textcolor{jsonkey}{"scenario"}@*): "S2",
  (*@\textcolor{jsonkey}{"authoring\_difficulty"}@*): 1,
  (*@\textcolor{jsonkey}{"datasets"}@*): ["H3K27ac-resgen.multivec"]
}
\end{lstlisting}
\end{minipage}\end{lrbox}

\begin{figure}[h]
\centering
\includegraphics[width=0.8\columnwidth]{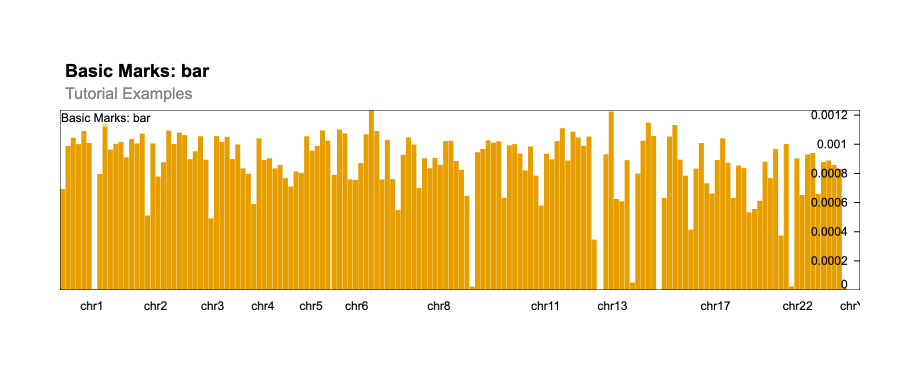}\\[2pt]
\vspace{-5mm}
\resizebox{0.8\columnwidth}{!}{\usebox{\datasetcodebox}}
\vspace{-3mm}
\caption{Example dataset entry. \texttt{query} is the raw user utterance used as input to the LLM. \texttt{Q} and \texttt{datasets} are reference fields used for evaluation: \texttt{Q} provides a disambiguated ground truth intent (S1), and \texttt{datasets} lists the expected data sources.}
\label{fig:dataset-example}
\end{figure}

\begin{figure*}[!t]
    \centering
    \includegraphics[width=1\textwidth, alt={}]{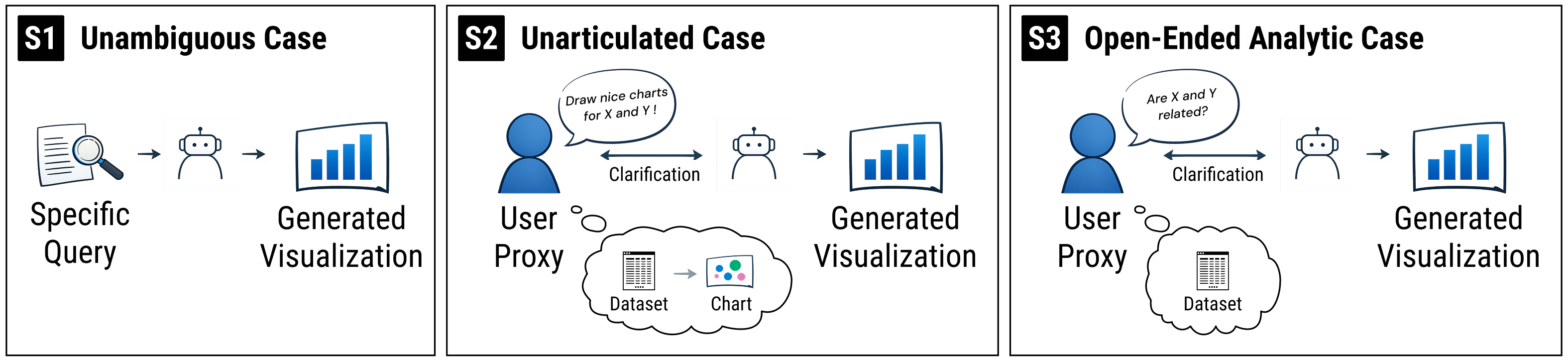}
   
    \caption{%
    Three query scenarios used in the evaluation. (S1) The query fully specifies the intended visualization; no user interaction is required. (S2) The user has a target dataset and visualization in mind but expresses only a vague request; a UserProxy answers clarification questions. (S3) The user poses an analytical goal without prescribing a visual form, requiring the system to make domain-informed design decisions.
    }
    \label{fig:eval-scenarios}
         \vspace{-3mm}
\end{figure*}

\textbf{Scenarios}.
Natural language queries can be phrased as a question or a command, and the amount of explicit information about chart types and data attributes varies~\cite{srinivasanCollectingCharacterizingNatural2021}.
In data-intensive domains such as genomics, queries often blend analytical goals with domain-specific terminology. This gives rise to three scenarios (S1--S3). S1 queries are unambiguous in both visualization intent and data attributes. Although users rarely provide this idealized input, they serve as a useful baseline. S2 and S3 queries are increasingly ambiguous. S2 is incomplete in data references (e.g., metadata-based rather than explicit file names) or visualization intent, while S3 frames requests in terms of analytical goals that imply the visualizations without naming them. From another perspective, drawing on visualization task taxonomies~\cite{brehmerMultiLevelTypologyAbstract2013}, S1 and S2 can be seen as \emph{how} tasks in visualization design, while S3 expresses a \emph{why} task~(\cref{fig:eval-scenarios}). 

\textbf{Complexity Levels}. Genomic visualization specifications vary considerably in authoring complexity, from single-track charts to interactive genome browsers. %
We define three levels (L1--L3) based on the structural and data-handling demands of the specification.
L1 (basic) covers single-track or simple multi-track charts using standard mark types and simple data sources that do not require complex transformations.
L2 (intermediate) introduces one or more authoring challenges, such as brushing and overview+detail interactions, semantic zoom, or specialized data types such as BAM and VCF. L3 (complex) consists of application-style genomic browser specifications that compose multiple coordinated views, heterogeneous data types, and domain-specific annotation layers, having more tracks, views and complex arrangement patterns. These levels were defined based on expert knowledge of the genomics visualization domain, but are not intended as a definitive complexity taxonomy.

\subsection{Method}
 To identify where the vanilla LLM succeeds and fails, the first author manually inspected a representative sample (45\%) of triplets, rendering each spec and comparing it against the ground truth focusing on three key questions: how well does the visualization represent the ground truth (S1), how reasonable is the visualization given a
vague description (S2), and how adequate is the visualization for the analytical goal (S3). Note excerpts were iteratively grouped and abstracted into eight quality dimensions. 
These dimensions were then developed into a scoring rubric for LLM-as-a-judge evaluation of all 159 outputs (Sec.~\ref{sec:evaluation}).

For S2 and S3, where queries likely require user clarification or interpretation, a user proxy agent, prompted with the ground truth specification and dataset metadata, responded to agent clarification requests on behalf of the user~(Fig.~\ref{fig:eval-scenarios}).

\subsection{Eight Quality Dimensions}
\begin{figure*}[!t]
    \centering
    \includegraphics[width=1\textwidth, alt={}]{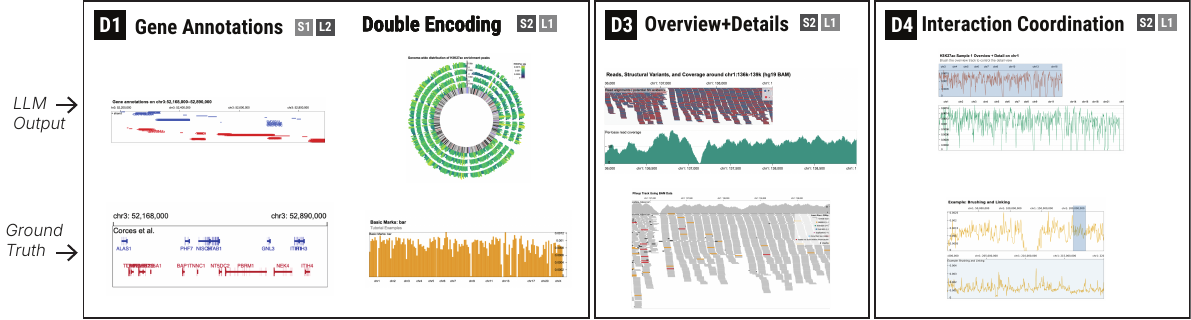}
    \caption{Representative failure cases from the vanilla LLM baseline. (D1) gene annotation tracks lacking directionality and exon structure (left), and double-encoding of the same quantitative variable on both height and color (right). (D3): detail view rendered before the overview, reversing the expected reading order. (D4) brush range wrong, breaking the overview+detail interaction.}
    \label{fig:failure-cases}
      \vspace{-3mm}
\end{figure*}

We identified eight recurring themes, each describing a distinct aspect of visualization output. To enable systematic scoring for the LLM-as-a-judge (Sec.~\ref{sec:evaluation}), we translated each theme into an evaluation criterion with a directional framing. Below we present a summary of the main observations; Fig.~\ref{fig:failure-cases} illustrates representative failure examples.

\textbf{D1: Mark \& Encoding Appropriateness.} 
At L1 and in S1, mark choices were generally appropriate --- \code{bar} for quantitative data, \code{area} for signal profiles, 
\code{rect} for intervals. As complexity increased, three persistent weaknesses emerged. First, conventional genomic visualization idioms for gene annotation and cytogenetic band tracks were a consistent weak spot. The model struggled to produce composite glyphs these idioms require: gene tracks rarely showed directionality, exon boundaries, or readable labels, and cytoband tracks lacked centromere triangles and visibility-filtered text. Second, structural variant encoding was fragile at L2--L3, with \code{withinLink} or \code{betweenLink} marks frequently missing colored strokes or having incorrect start/end field mappings. Third, the model frequently double-encoded the same quantitative variable on both bar height and color, particularly in response to ambiguous queries (S2). This suggests a tendency to over-specify encodings when the prompt is vague, rather than defaulting to a simpler, conventional choice, a pattern that may reflect LLM behavior of attempting to over-satisfy under uncertainty. 

\textbf{D2: Query Compliance.}
Query compliance was mostly assessed for S1, and here L1 visualizations were largely compliant. For higher complexity levels, we observed some cases where the requested genomic region does not match. Generally at L2 and L3, there are more missing tracks or views, caused both by spec errors that prevented rendering and by omission of requested components such as data stratification or filtering. We note that for S2 and S3, the boundary between ``incomplete'' and ``acceptably different interpretation'' was less clear, suggesting that this criterion is most informative for S1 queries.

\textbf{D3: Layout \& Composition.}
In S3, layout and composition issues were most common, particularly in the spatial organization of views and tracks for effective comparison and reading, as queries often prompted complex multi-view outputs. The most common problem was failure to \code{overlay}: tracks that should share a single view (e.g., cytobands over a reference track, text labels on gene bodies) were instead placed as separate tracks or views, wasting screen space and hindering comparison. In S1, view ordering was occasionally reversed from the ground truth, with detail appearing before overview, and related tracks were sometimes misaligned. Interestingly, the model also chose defensible alternative arrangements, e.g., \code{serial} arrangement for two genomic regions, or \code{linear} instead of \code{circular} layout for whole genome overview. The overall pattern suggests that the model treats each track as an independent unit rather than reasoning about the spatial relationships between them, showing a tendency to separate views that a domain expert would merge.

\textbf{D4: Interaction \& Coordination.}
Brush implementation was the single most consistent failure across the dataset: the \code{brush} was absent or non-functional in the vast majority of cases. Coordinated panning and zooming, dependent on a matching \code{linkingId} across views, showed inconsistent behavior, succeeding in some specifications but failing in structurally similar ones with no clear predictor of success. More complex semantic zooming in L2 and L3 specifications was also largely absent. As a result, in canonical visualizations, cytoband labels, nucleotide labels, and gene names were either always visible (causing overplotting) or never visible. This suggests that correctly linking views might be a higher level task that the vanilla LLM struggles with.

\textbf{D5: Axes \& Legends.}
This was a common issue across most scenarios and levels, but was concentrated at L2 and L3 in S2, where multi-track views lacked the reference elements needed to orient the reader. The most striking pattern was a bias toward suppressing genomic axes (often the x-axis), whereas quantitative axes (typically the y-axis) were generated more reliably, sometimes redundantly, suggesting that the model treats quantitative axes as more essential than positional ones. Also titles as reference elements were often omitted, making complex multi-track or view visualizations essentially unreadable without inspecting the underlying spec. This points to a gap between generating content and generating the readability frames and metadata needed to interpret it.

\textbf{D6: Proactive Enhancements.}
This dimension is predominantly relevant for S2 and S3, where the open-ended prompt leaves room for the LLM to add context beyond what was explicitly requested. Very common additions were a \code{tooltip} and extra genomic tracks such as cytobands or gene annotations. It seems like the model chose tooltips as a substitute for axes and legends rather than a complement. Extra cytoband and gene tracks were conceptually reasonable for providing navigational context. However, they suffered from the same rendering failures observed in D1 and additionally showed awkward placement, for example between data tracks that were clearly grouped. Some additions were also highly uninformative, such as rendering text labels for variables that had the same value across all data points in a dataset. The pattern suggests that the model might have some intuition that genomic visualizations should include context tracks, but applies this without evaluating whether the specific addition is truly informative or well-placed.

\textbf{D7: Styling \& Visual Clarity.}
This was the smallest theme but still pointed to flaws affecting exploration and readability. For example, the model occasionally diverged from Gosling’s built-in default colors without clear rationale, instead choosing less perceptually distinguishable alternatives. We also observed cases where different colors were assigned to the same data variable across linked views, hindering cross-view consistency. Notably, Gosling does not currently enforce cross-view color consistency at the grammar level, so an LLM capable of reasoning about this could provide added value beyond what the specification language itself supports.

\textbf{D8: Data Compliance.}
Data compliance was consistently reliable for matching the required source files, even in S2 and S3 where the model had to infer the correct source from multiple entries using the provided metadata and catalog. Failures were concentrated at L2 and L3, where multi-field datasets and complex transformations were required. The primary failure mode was field parsing: tracks appeared to be empty not because data was absent but because field names were incorrectly mapped. In several cases, the data was technically present (as observed in the spec) but the rendering failure made it invisible. 
We note that data compliance is often entangled with D1 and D2: a missing track can reflect a field parsing error (D8), an encoding failure (D1), or an omission of a requested query component (D2).

\section{Agentic Authoring}
\label{sec:agentic-authoring}

The eight quality dimensions identified in Sec.~\ref{sec:compare-vanilla-llm} reveal some common failure modes that motivate three design decisions for an agentic authoring system:
\begin{enumerate}
    \item~\textbf{Agentic interaction.} Failures across the eight dimensions are often interdependent: a missing track may reflect an encoding failure (D1) or a query omission (D2). %
    They are also not always immediately apparent from the specification alone: a missing \code{linkingId} (D4) is technically visible in the JSON specification, but requires reasoning across views, while mark-type errors (D1) are often more obvious in the rendered screenshot. Different failure types require different inspection methods. The system must therefore support an iterative \emph{generate–render–inspect–fix} loop rather than relying on a single generation pass with only spec-level validation. 
    \item~\textbf{Multi-agent specialization.}\label{decision-2} The eight dimensions split into domain-specific groups that require distinct expertise and tooling. Data wiring (D2, D8) depends on data catalog lookup and field inspection; visual encoding (D1) requires knowledge of Gosling mark semantics; spatial structure (D3, D4) involves view arrangement and linking logic. Dedicating an agent to each group lets it experiment within its scope without destabilizing other concerns (i.e., separation of concerns).
    \item~\textbf{Reviewer.}\label{decision-3} A reviewer evaluates the full output, with particular oversight on dimensions requiring broader visualization literacy beyond genomics-specific concerns, including axes and legends (D5), proactive enhancements (D6), and styling (D7), such as title presence, legend consistency, and cross-view color consistency. This separate reviewer can flag issues across these dimensions, and modify the specification where needed.
\end{enumerate}

\begin{figure*}[!t]
    \centering
    \includegraphics[width=0.85\textwidth, alt={}]{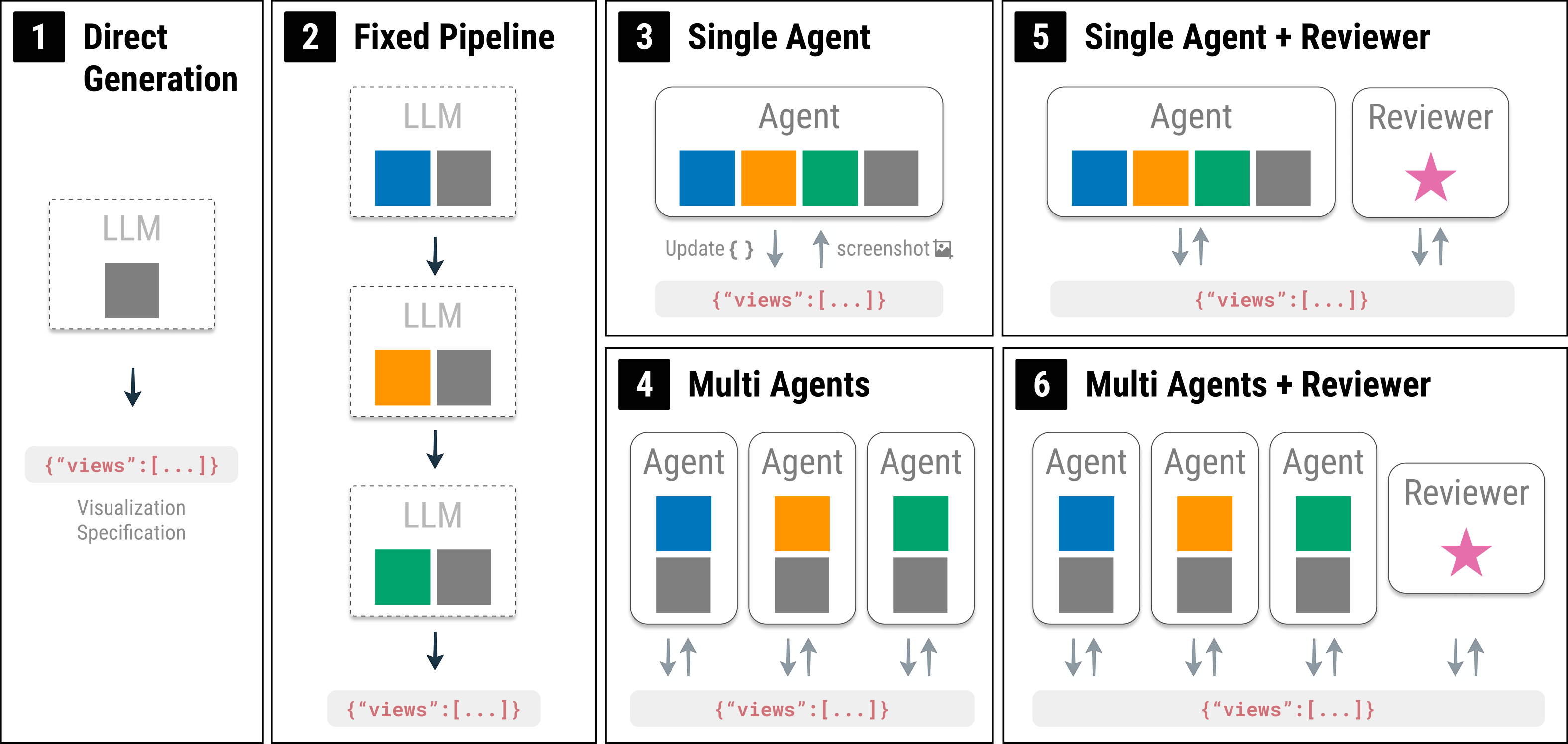}
    \caption{%
Six authoring schemes used in the evaluation, ordered by increasing structure and specialization.
Each colored block indicates its areas of responsibility:
query-data alignment (\csq{cQDA}),
visual encoding (\csq{cVEC}),
layout and structure (\csq{cLS}),
and shared context such as visualization specification grammar.
(1)~Direct Generation produces a complete spec in a single LLM call.
(2)~Fixed Pipeline decomposes the authoring process into three LLM steps, where the second step is invoked once per view before the third assembles the full specification.
(3)~Single Agent uses one iterative agent that handles all responsibilities in a screenshot-driven review loop.
(4)~Multi Agents splits these responsibilities across three specialist agents who share a single spec and take turns reviewing.
Schemes~5 and~6 extend schemes~3 and~4, respectively, with a Reviewer that performs a holistic quality review using domain knowledge bases~(\csq{cPG}).}
    \label{fig:schemes}
      \vspace{-3mm}
\end{figure*}

\subsection{Six Authoring Schemes}
\label{sec:schemes}

We define six authoring schemes that progressively introduce structure, iteration, and specialization~(\cref{fig:schemes}).

\textbf{Non-agentic baselines.}
\textit{Direct Generation (\cref{fig:schemes}.1):} a single LLM call receives the natural language query, data catalog, and Gosling documentation, and outputs a complete specification. This setup does not involve any tools or iterations. This scheme corresponds to the vanilla LLM baseline introduced in \cref{sec:compare-vanilla-llm}. \textit{Fixed-pipeline (\cref{fig:schemes}.2):} a hand-crafted three-step chain that decomposes the authoring task into sub-tasks, following the general visualization reference model by Card et al.~\cite{card2009information}~(i.e., data, visual encodings, view structures). This decomposition is inspired by the six-step pipeline of GenoREC~\cite{pandeyGenoRECRecommendationSystem2023a}, a recommendation system for genomics visualization, but consolidates those steps into three. We observed in the evaluation of the vanilla LLM that dimensions are often intertwined (e.g., D1 and D8, D3 and D4), so excessive fragmentation risks losing dependencies between steps that the LLM needs to reason about. Step 1 (data resolution) selects datasets and fields from the catalog; Step 2 (encoding) generates track-level specs with marks and visual channels. Finally, Step 3 (composition) assembles tracks into the final Gosling spec. Each step is an LLM call with structured output, and Step 2 can make \texttt{T} calls depending on the number of tracks. Similar to direct generation, this scheme has no tool-calling loop or rendered-output feedback, but Steps 2 and 3 each allow up to two validation retries.

\textbf{Agentic schemes (2×2).} Design decisions \#\ref{decision-2} and \#\ref{decision-3} define two independent factors: single vs.\ multi-agent authoring, and presence of a reviewer, resulting in four agentic schemes in total. \textit{Single-Agent~(\cref{fig:schemes}.3):} one generalist agent (\texttt{GoslingAuthor}) handles all aspects of building the specification, including query and data alignment (\csq{cQDA}), visual element compliance (\csq{cVEC}), layout and structure (\csq{cLS}), and presentation decisions. \textit{Multi-Agents~(\cref{fig:schemes}.4):} three specialist agents split the task. A Query-Data Alignment agent (\csq{cQDA}~\texttt{QDA}) manages datasets, fields, and transforms (D2, D8). A Visual Element Compliance agent (\csq{cVEC}~\texttt{VEC}) handles marks, encodings, and overlays (D1). A Layout and Structure agent (\csq{cLS}~\texttt{L\&S}) manages view arrangement, linking, and dimensions (D3, D4). \textit{+Reviewer~\csq{cPG}} (applicable to both, \cref{fig:schemes}.5--6): a holistic reviewer that performs a final quality pass over the complete specification by re-validating all dimensions, with particular attention to presentation (axes, legends, styling; D5, D7) and contextual enhancements (D6), and editing the specification if needed.
The core domain guidance (given in the modules QDA, VEC, and L\&S) is consistent across all schemes except the Direct Generation baseline.

\subsection{Shared Agentic Architecture}
\label{sec:shared-architecture}

All four agentic schemes share the same coordination protocol. A coordinator routes control to each agent in sequence. Each agent reviews the current spec and screenshot, then either calls \texttt{update\_spec} to modify the specification or returns \texttt{confirm}. Any spec update triggers a new screenshot render and restarts the review cycle from the updating agent, requiring all reviewers to validate the updated specification again. The loop terminates when all agents confirm or a maximum of 15 rounds is reached. Agents are further scoped by their available tools: the QDA agent has access to the data catalog and data inspection tools. VEC and L\&S agents can only read and update the spec. The Reviewer additionally has access to the data catalog and data inspection tools to verify field references and filter values during its quality pass.

\subsection{Visualization Knowledge Bases}
\label{sec:kb}
Some design judgments, such as whether an encoding is effective, color assignments remain consistent across views, or layout and arrangement choices support comparison, go beyond what the Gosling grammar itself enforces. We therefore curated a visualization design knowledge base for the reviewer agent, in three layers of guidance. For general visualization principles (1), we incorporated design guidelines from VisGuides~\cite{diehl2018visguides}, the subset of focused design feedback questions filtered by Kim et al.~\cite{kimHowGoodChatGPT2025}, the 23 design issues across expressiveness, effectiveness, interpretability, legibility, and perception identified by BaviSitter~\cite{choi2024bavisitter}, and the multi-view dashboard quality criteria from Waltzboard~\cite{choi2025waltzboard}. For formalized design constraints (2), we drew on Draco's ASP-based knowledge representation~\cite{moritz2018formalizing, yang2023draco}, which enables scalable encoding of soft and hard visualization design rules. Lastly, we add genomics-specific guidance (3) by incorporating the task taxonomy by Nusrat et al.~\cite{nusratTasksTechniquesTools2019}, the recommendation rules from GenoREC~\cite{pandeyGenoRECRecommendationSystem2023a}, and automatically synthesized Gosling-specific design rules~\cite{kimAutomaticSynthesisVisualization2026}.

This layered approach addresses a known gap between visualization research and practice~\cite{ahnUnderstandingWhyChatGPT2025}: design knowledge is scattered across papers, codebases, and forums, and perception-based findings can be conflicting~\cite{zeng2023too}. By consolidating these sources into structured prompts, following Kim et al.'s template for actionable design guidelines ``\emph{When/if {Context}, {Approach}, because of {Problem}, for {Purpose}}''~\cite{kim2025understanding}, we give the reviewer agent access to actionable design knowledge at inference time without requiring fine-tuning.

\section{Evaluation}
\label{sec:evaluation}

\subsection{Experimental Design}
\label{sec:experimental design}

We evaluated all six schemes on the 53 ground-truth Gosling specifications introduced in Sec.~3. Each specification belonged to one of three complexity levels (L1--L3) and was evaluated under three query scenarios (S1--S3), resulting in 159 cases in total. Each case was run three times per scheme, yielding 477 runs per scheme and 2,862 runs in total. All schemes used \textsc{gpt-5.4-2026-03-05} as the underlying language model. We address the following research questions:

\begin{itemize}
\item \textbf{RQ1:} How does agentic authoring compare to non-agentic baselines?
\item \textbf{RQ2:} How do architectural choices within agentic schemes affect output quality?
\item \textbf{RQ3:} How do task characteristics such as specification complexity and query ambiguity affect scheme performance?
\end{itemize}

\subsection{Method}

We assess output quality along two complementary axes. \textbf{Perceived quality} is scored by an LLM-as-a-judge pipeline on the eight quality dimensions from \cref{sec:compare-vanilla-llm} (D1--D8), each rated on a 1--5 Likert scale (5 = best); the composite is the mean of all applicable dimensions per observation. \textbf{Structural similarity} is measured by CFG similarity~\cite{nguyenGeraniumMultimodalRetrieval2025}, which compares generated and reference specifications via context-free grammar decomposition (0--1 scale). The two can diverge: an output may be judged as a good visualization while deviating structurally from the reference.

To identify which factors most strongly influence output quality among the many possible comparisons across six schemes, three scenarios, and three complexity levels, we fit linear mixed-effects models addressing the three research questions (RQ1--RQ3), with specification as a random intercept to control for difficulty differences. We group the six schemes into three \emph{scheme types}: direct generation, fixed pipeline, and agentic (the four agent-based schemes pooled). We focus on marginal $R^2$ (how much variance the predictors explain) and $\Delta R^2$ (how much each individual factor contributes) to distinguish effects that are statistically significant from those that are also practically meaningful. We report fixed-effect coefficients ($B$).

\subsection{Results}

\subsubsection{Agentic schemes outperform baselines on perceived quality but not structural similarity.}

\begin{figure}[h]
  \centering
  \includegraphics[width=\columnwidth]{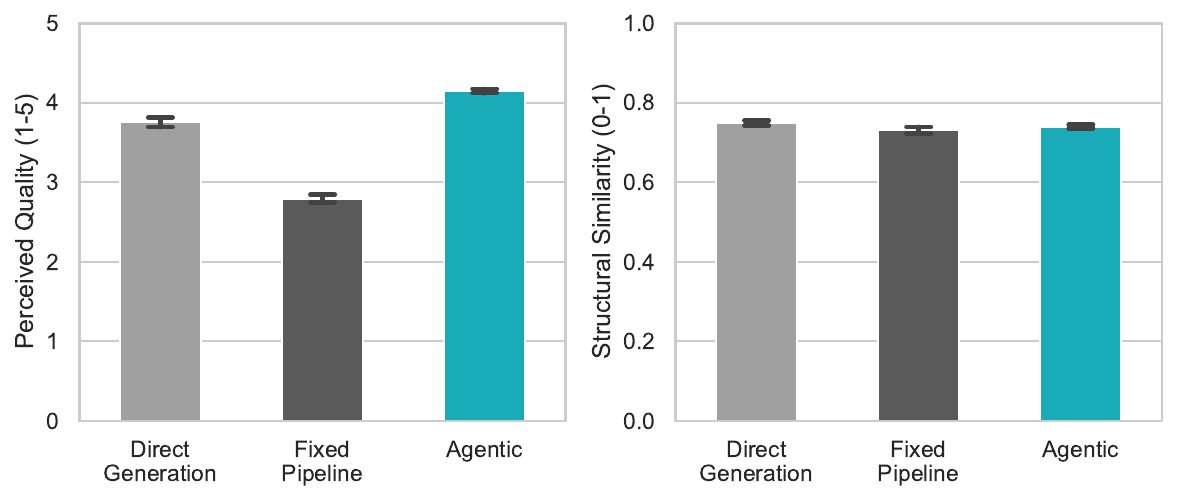}
  \caption{Perceived quality (left) and structural similarity (right) by scheme type. Error bars show 95\% CI. Agentic schemes outperform both baselines on perceived quality, while structural similarity remains comparable across scheme types.}
  \label{fig:tier-main}
  \vspace{-3mm}
\end{figure}

Agentic schemes produced significantly higher perceived quality than direct generation and fixed-pipeline baselines, with scheme type explaining 43.2\% of variance (agentic vs.\ direct generation: $B = +0.199$, $p < .001$; fixed pipeline vs.\ direct generation: $B = -1.222$, $p < .001$). Scheme type had minimal impact on structural similarity ($\Delta R^2 = 1.3\%$) despite statistically significant pairwise differences (agentic vs.\ direct generation: $B = -0.045$, $p < .001$; fixed pipeline vs.\ direct generation: $B = -0.036$, $p < .001$). \cref{fig:tier-main} shows both measures side by side.

\subsubsection{Agentic schemes may be more resilient to specification complexity.}

\begin{figure}[h]
  \centering
  \includegraphics[width=\columnwidth]{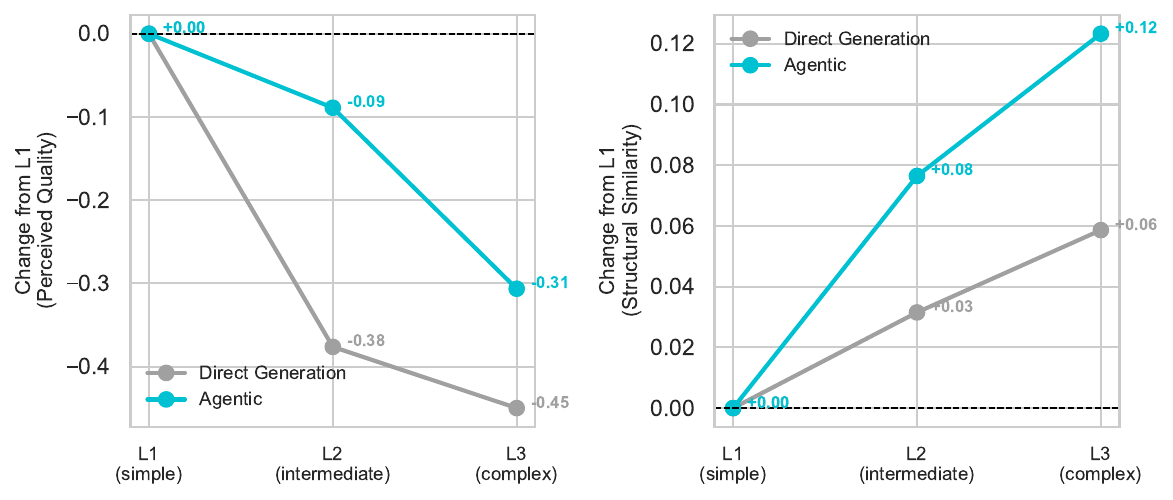}
  \caption{Change relative to L1 for each scheme type. Left: perceived quality; right: structural similarity. Both scheme types degrade with increasing complexity, although the strongest effects occur for Direct Generation. For structural similarity, agentic schemes close and eventually reverse the gap at L3.}
  \label{fig:drop-from-l1}
    \vspace{-3mm}
\end{figure}

Specification complexity significantly interacted with scheme type for both measures (\cref{fig:drop-from-l1}). For perceived quality, direct generation quality dropped $-0.38$ from L1 to L2 and $-0.45$ to L3, while agentic schemes dropped only $-0.09$ and $-0.31$ respectively (scheme type $\times$ L2: $B = +0.286$, $p < .001$; scheme type $\times$ L3: $B = +0.148$, $p = .037$). For structural similarity, the mixed-effects model estimated agentic schemes to score lower than direct generation at L1 ($B = -0.045$), but to close the gap at L2 ($B = +0.045$, $p < .001$) and outperform it at L3 ($B = +0.065$, $p < .001$). Query ambiguity, in contrast, did not moderate the agentic advantage for either measure ($\Delta R^2 \leq 0.1\%$ for both). While these interactions are statistically significant, they explain only $0.8\%$ and $1.0\%$ of variance respectively, and should be interpreted as suggestive rather than definitive.

\subsubsection{Using multiple agents or adding a reviewer does not improve quality.}

Within the four agentic schemes, neither using multiple specialist agents nor adding a reviewer significantly affected perceived quality ($p = .850$ and $p = .488$ respectively), and the full model explained only 6.6\% of variance (marginal $R^2 = .066$), almost entirely from ambiguity and complexity rather than architecture choices. For structural similarity, both factors were statistically significant but explained less than 1\% of variance combined ($\Delta R^2 = 0.2\%$ and $0.4\%$). Within this negligible range, using multiple agents produced slightly higher structural similarity ($B = +0.014$, $p = .003$) but slightly lower perceived quality ($B = -0.009$, n.s.), while adding a reviewer lowered structural similarity ($B = -0.016$, $p < .001$) without any compensating improvement in perceived quality.

\subsubsection{Reviewer and layout agents contribute less.}

\begin{figure}[h]
  \centering
  \includegraphics[width=\columnwidth]{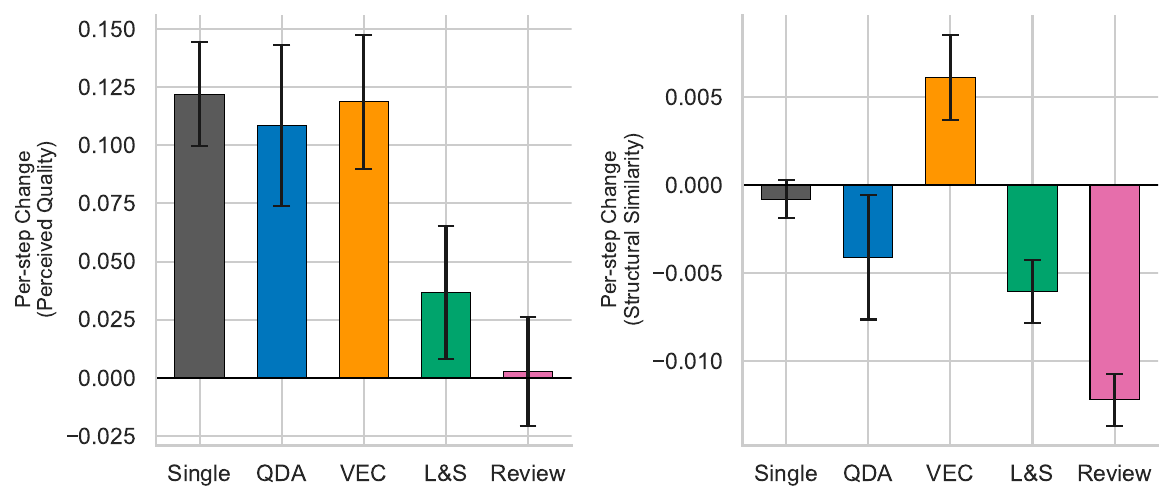}
  \caption{Mean per-step change in perceived quality (left) and structural similarity (right) by agent type. Error bars show 95\% CI. Authoring agents contribute more than reviewers or layout agents.}
  \label{fig:agent-deltas}
\end{figure}

To understand why additional agents do not improve overall quality, we examined per-step quality changes across 6,663 refinement steps (\cref{fig:agent-deltas}). Only three out of five agents---Single, QDA, and VEC---improved perceived quality by a comparable amount per step ($B \approx +0.12$, all $p < .001$; no significant differences among them). L\&S similarly underperformed on perceived quality ($B = -0.086$ vs.\ Single, $p < .001$). The Reviewer, although empowered to edit the spec, is positioned as a final-pass reviewer rather than a primary author. It contributed near-zero improvement in perceived quality ($B = -0.119$ vs.\ Single, $p < .001$) while producing the largest degradation in structural similarity ($B = -0.011$, $p < .001$).  Notably, VEC was the only agent that improved structural similarity per step ($B = +0.007$, $p < .001$). These per-step effects are small ($\Delta R^2 = 0.8\%$ for perceived quality, $2.4\%$ for structural similarity) and confounded with pipeline position, but they are consistent with the scheme-level finding that adding a reviewer does not help.

\subsubsection{Agentic schemes come at significantly higher cost.}
Agentic schemes require substantially more time and tokens than Direct Generation. Adding a reviewer has a negligible effect on turns in the single-agent setting (4.4 $\rightarrow$ 4.4 on average; $+0.02$ unrounded) but a clear effect with multiple agents (4.5 $\rightarrow$ 6.0), and in both cases it raises cost (\$3.28 $\rightarrow$
 \$3.49; \$3.65 $\rightarrow$ \$4.98 per run). Moving from a single agent to multiple agents increases API calls (13.8 $\rightarrow$ 19.8) and cost regardless of whether a reviewer is included. Among agentic schemes, the simplest configuration --- Single Agent at \$3.28/run --- achieves comparable quality at the lowest cost.

\subsubsection{Perceived quality and structural similarity capture independent aspects of output quality.}

\begin{figure}[t]
  \centering
  \includegraphics[width=0.7\columnwidth]{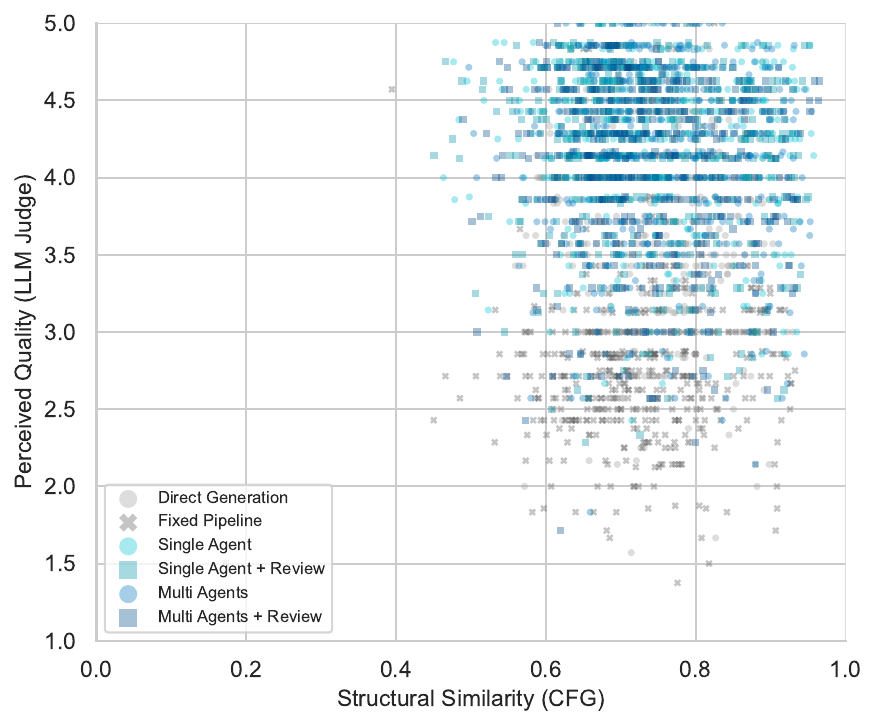}
  \caption{Perceived quality vs.\ structural similarity for all observations, colored by scheme. The two measures are independent ($r = 0.001$), with outputs distributed evenly across both axes regardless of scheme.}
  \label{fig:cfg-vs-judge}
\end{figure}

\cref{fig:cfg-vs-judge} plots the two measures against each other for 2,843 observations. They are effectively independent ($r = 0.001$, $p = .978$), both overall and within each scheme. Outputs are distributed evenly across both axes, confirming that the two metrics capture different aspects of visualization quality.

\subsection{Illustrative Example: Single-Cell Epigenomics Dataset}

To ground the quantitative findings above in a concrete case, we illustrate scheme behavior on a complex L3 specification based on the Corces et al. single-cell epigenomics dataset~\cite{corcesSinglecellEpigenomicAnalyses2020}, which requires a cytoband overview linked via brush interaction to multiple bigwig signal tracks and structural variant annotations. We show the S2 case in \cref{fig:teaser}; corresponding figures for S1 and S3 are in the supplementary material.

In S1 (ground truth reproduction), all schemes broadly succeeded in reproducing the requested multi-track layout with the correct data mappings and signal encodings. The main point of failure was consistent across schemes: no scheme rendered the cytoband track correctly, though agentic schemes came closer to a recognizable cytoband than the baselines. Agentic schemes also most consistently generated the brush mark for overview-to-detail linking, though the brush did not always function correctly.

In S2 (ambiguous query), agentic schemes more clearly separated from the baselines. They produced more expressive encodings and were more likely to include the brush interaction for coordinated exploration. The overall visual quality and completeness of agentic outputs exceeded those of direct generation, suggesting that iterative refinement against the rendered output helps the model make better design choices when the prompt leaves room for interpretation.

In S3 (data-driven question), an unexpected instability emerged: linking behavior between views became unreliable across all agentic schemes, with coordinated navigation breaking down in ways not observed in S1 or S2. The brush, present in earlier scenarios, was now omitted. This regression suggests that when the model must simultaneously reason about an analytical question and compose a multi-view visualization, interaction features are among the first things to be dropped. This is consistent with our earlier finding on the interaction \& coordination dimension (D4), where linking emerged as a weak area.

\begin{figure*}[t]
  \centering
  \includegraphics[width=\linewidth]{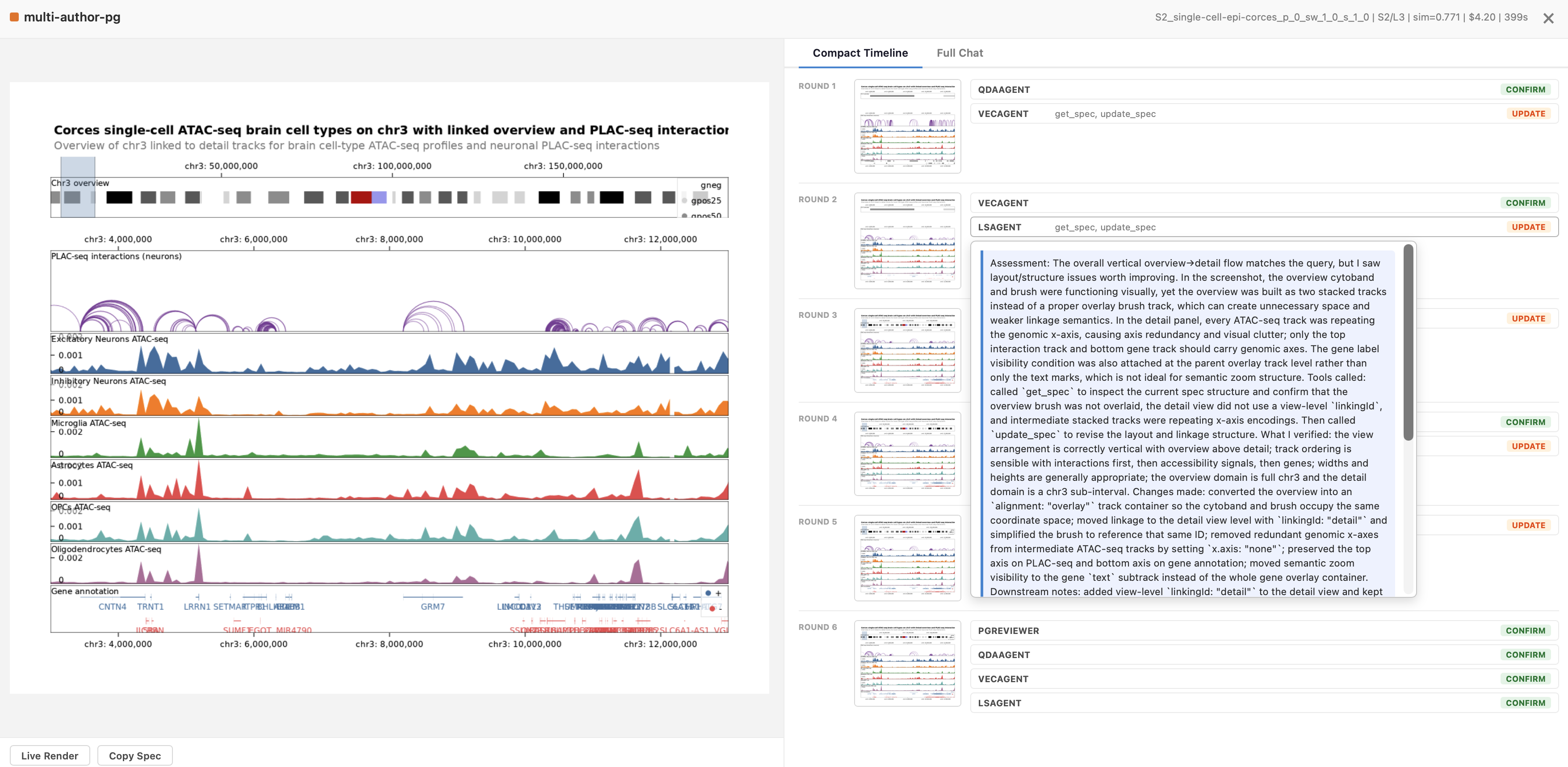}
  \caption{Multi Agent + Reviewer scheme (S2/L3): iterative refinement of the Corces et  
  al. single-cell epigenomics specification over six rounds, showing agent actions
   and intermediate renders.}
  \label{fig:corces-iteration}
\end{figure*}

The iterative nature of the agentic schemes is visible in the step-by-step evolution of individual runs. For instance, in the Multi Agents + Reviewer scheme~(\cref{fig:corces-iteration}), the QDA agent first corrected a data transformation that had been misapplied in earlier iterations, and the Reviewer subsequently added a cytoband track for navigational context. This incremental layering of fixes and additions illustrates how the separation of concerns across agents can lead to cumulative improvements that non-agentic baselines did not produce.

\begin{figure}[t]
  \centering
  \includegraphics[width=0.8\columnwidth]{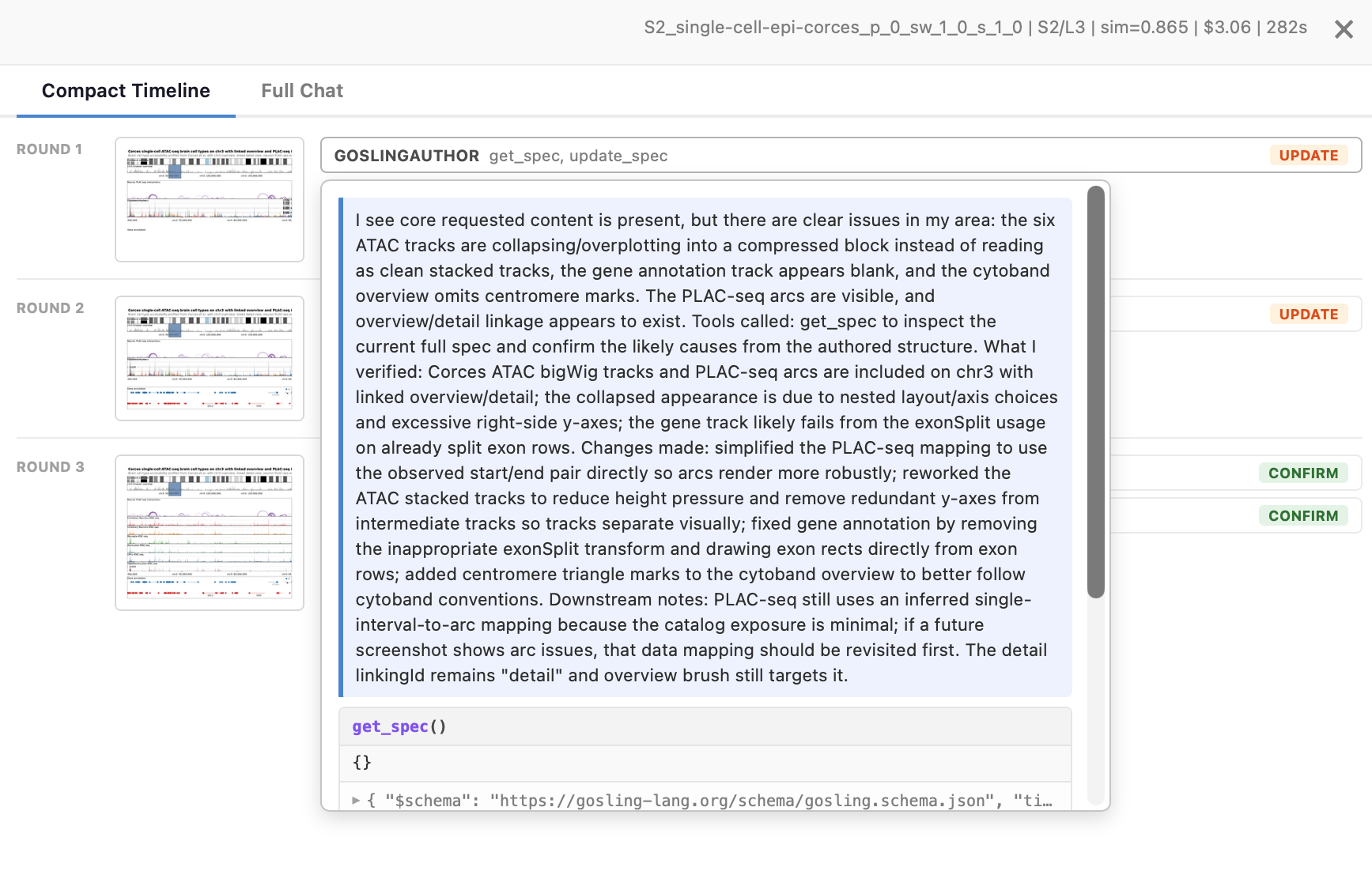}
  \caption{Single Agent + Reviewer scheme (S2/L3): iterative refinement of the Corces et  
  al. single-cell epigenomics specification over three rounds, showing agent actions
   and intermediate renders.}
  \label{fig:corces-sa}
  \vspace{-5mm}
\end{figure}

A run from the Single Agent + Reviewer scheme shows another interesting progression (\cref{fig:corces-sa}). The generalist author initially produced gene annotations with the typical failures also seen in Direct Generation: no directionality, missing labels, overplotted text. Over successive iterations, the author refined the gene track to correctly show strand direction and readable labels, and changed an overlaid line encoding to a stratified arrangement to improve comparison effectiveness. The Reviewer performed a final quality pass but did not substantially alter the output, suggesting that a capable generalist agent can self-correct on canonical genomic encodings through iterative refinement alone.

\section{Discussion}
\label{sec:discussion}

\subsection{Flexible Agents over Rigid Grammars}

Agentic schemes were the strongest predictor of perceived quality in our evaluation, outperforming both Direct Generation and Fixed Pipeline. This advantage was consistent across complexity levels, with a small but significant interaction suggesting that agentic schemes degrade less than Direct Generation as specifications grow more complex.

Surprisingly, the Fixed Pipeline---a hand-crafted three-step decomposition with dedicated prompts and domain-specific guidelines---performed worse than even a single unconstrained LLM call given only the grammar documentation. This is notable because a structured pipeline is arguably the most natural next step after Direct Generation, the approach a practitioner would try first when one-shot results fall short. We invested substantial effort in its design, iterating on prompt engineering for data resolution, encoding selection, and view composition, yet many small failures persisted: incorrect field mappings, misapplied genomics conventions, broken interactions between views. These issues were especially pronounced for specifications involving brushing, overview-detail coordination, or semantic zoom, where a single change can cascade across views and encodings. %
Further prompt engineering did not resolve this error accumulation.

Agentic iteration is novel in that it gives the system the ability to observe its own output and self-correct. In our experiments, this took the form of a generate-render-inspect-fix loop where agents iteratively revised their specifications based on rendered screenshots, sidestepping the need to get everything right on the first pass. This approach is particularly valuable for specialized visualization grammars like Gosling. Such grammars are carefully crafted to serve expert communities, shaped by domain-specific conventions and accumulated design decisions. This makes them expressive but also heavy: they evolve slowly, updates are costly, and their learning curve is steep. Agentic iteration offers a new direction for working within these constraints without requiring changes to the grammar itself. Beyond enabling correction, the rigidity of a specialized grammar may actually help the agent; even when a model has access to the grammar, it does not necessarily use it well. 
As our analysis of vanilla LLM generation showed (\cref{sec:compare-vanilla-llm}), even with the Gosling grammar constraining what can be expressed, a single LLM call tends toward hallucination, over-specification, and stylistic excess. The grammar and the agent complement each other by design: a grammar's structured output acts as a guardrail, limiting the space of valid outputs, while iterative refinement against the rendered output lets the agent flexibly search within that space and correct errors a single pass leaves behind.

\subsection{More Complex Architectures Require More Precise Control}

Using multiple specialist agents or adding a reviewer did not meaningfully improve output quality. A single agent achieved the same perceived quality as more complex configurations at the lowest cost.

Our per-step analysis offers some insight into why. Although all agents could edit the specification, only three agents that handle primary authoring (QDA, VEC, and Single) contributed comparable per-step improvements, while those operating at a higher level of abstraction contributed less. L\&S, responsible for layout and coordination, showed smaller per-step improvements, and the reviewer showed the least contribution of all. These differences were small in magnitude, but they suggest that simply adding more agents to a complex visualization authoring task is not inherently beneficial.

This suggests the benefit of agentic authoring lies not in how work is divided but in how tightly the agent's feedback mechanism is coupled to the grammar. Currently, agents rely on rendered screenshots, which provide holistic but coarse feedback. Finer-grained tools, such as per-track validation, sub-view rendering, or interaction-level debugging, could let agents inspect and correct more precisely. For grammars where syntactic correctness does not guarantee semantic validity, such tools would likely deliver larger gains than adding agents. Similarly, although the Reviewer could modify the specification, it rarely chose to. Future designs might scaffold the reviewer role to provide more structured feedback tied to specific grammar elements and act on the issues it identifies, rather than defaulting to passive critique.

\subsection{Augmenting Benchmarks with Agentic Evaluation}

Evaluation of LLM-based visualization authoring has largely relied on automated benchmarks, and this dependence is even stronger in specialized domains like genomics, where recruiting expert users is difficult and evaluation datasets are scarce. In this work, we attempted to go beyond static benchmarks by introducing a user proxy agent that simulates realistic clarification dialogues across three levels of query ambiguity. This allowed a single dataset of specifications to approximate how real users with varying levels of specificity might interact with the system.

The expected moderation by query ambiguity did not materialize in our results. This leaves open whether agentic schemes actually handle ambiguity well through their iterative process, or whether the proxy, which had access to ground truth metadata, did not faithfully simulate the difficulty of real user interactions. Simulating realistic user behavior within automated benchmarks remains a challenge, and distinguishing between these explanations will require evaluation with actual users in conversational settings.

\subsection{Limitations and Future Work}

\begin{itemize}
\item \textbf{Generalizability.} All experiments used \textsc{gpt-5.4-2026-03-05}; results may differ with other models. Our dataset draws from established Gosling examples that tend to use default options, potentially overestimating performance on novel designs.
\item \textbf{Evaluation.} Our evaluation was fully autonomous, with no human in the loop and no validation of the LLM judge against expert ratings. A conversational interface where users can steer and correct would reveal patterns that single-turn evaluation cannot. %
\item \textbf{Metrics.} Structural similarity and perceived quality are independent, raising the question of what constitutes a ``correct'' output when multiple valid visualizations exist. Future work could explore constraint satisfaction checking or user preference elicitation as alternatives to reference-based comparison.
\item \textbf{Agent design.} The Reviewer's ineffectiveness suggests redesign opportunities, such as scaffolding more decisive correction, providing structured feedback tied to grammar elements, or applying persona-inspired prompting. More broadly, agentic tools need tighter coupling with grammar semantics.
\item \textbf{Interaction modalities.} Supporting input beyond text, such as sketching or direct manipulation of the specification, could bridge the gap between user intent and system output.
\end{itemize}

\section{Conclusion}
\label{sec:conclusion}

We investigated how agentic LLM-based approaches can support the authoring of complex, domain-specific visualizations, using interactive multiview genomics visualization in Gosling as a testbed. Agentic schemes outperformed baselines, while more complex architectures did not yield further gains. Our findings suggest that specialized grammars and flexible agents complement each other, and that realizing this potential will require tighter tooling to bridge the two.

\section*{Supplemental Materials}
\label{sec:supplemental_materials}

All supplemental materials are available on OSF at \href{https://doi.org/10.17605/OSF.IO/UQE83}
{\texttt{osf\discretionary{}{.}{.}io\discretionary{/}{}{/}uqe83}}, released under a \href{https://creativecommons.org/licenses/by/4.0/}{\ccby{} CC BY 4.0 license}.
\acknowledgments{%
  The authors wish to thank Huyen N. Nguyen for helpful discussions about the Geranium dataset.
  This work was supported in part by grants from the National Institutes of Health (NIH R01HG011773, K99HG013348) and the Advanced Research Projects Agency for Health (ARPA-H AY2AX000028).
}

\bibliographystyle{abbrv-doi-hyperref}

\bibliography{references}

\end{document}